\documentclass[preprint, showpacs,preprintnumbers,amsmath,amssymb,nofootinbib]{revtex4}
\usepackage{mathrsfs}
\usepackage{amsmath}
\usepackage{amsfonts}
\usepackage{latexsym}
\usepackage{amsfonts}
\usepackage{graphicx}
\usepackage{epsf}
\usepackage{dcolumn}
\usepackage{bm}

\newcommand{\bea}[1]{\begin{eqnarray}\label{#1}}
\newcommand{\eea}{\end{eqnarray}}

\def\gsim{ \lower .75ex \hbox{$\sim$} \llap{\raise .27ex \hbox{$>$}} }
\def\lsim{ \lower .75ex \hbox{$\sim$} \llap{\raise .27ex \hbox{$<$}} }

\begin{document}
 \title{$f(T)$ models with  phantom divide line crossing}

\author{Puxun Wu
and  Hongwei Yu  }
\address
{  Center for Nonlinear Science and Department of Physics, Ningbo
University,  Ningbo, Zhejiang 315211, China
}

\begin{abstract}
In this paper, we propose two new models in $f(T)$ gravity to
realize the crossing of the phantom divide line  for the effective
equation of state, and we then study the observational constraints
on the model parameters.  The best fit results suggest that the
observations favor a crossing of the phantom divide line.

\end{abstract}
\maketitle

\section{Introduction}\label{sec1}
Various observations~\cite{Riess1998,Spergel2003,Tegmark2004,
Eisenstein2005} have confirmed the fact that our Universe is
undergoing an accelerating expansion and it entered this
accelerating phase only at the near past. The proposals that have
been put forth to explain this observed phenomenon can basically be
classified into two categories.  One is to assume the existence of
an exotic energy with negative pressure, named dark energy. The
simplest candidate of dark energy is the cosmological constant with
the equation of state $w=-1$~\cite{Weinberg1989}. It, however,
suffers from two serious theoretical problems, i.e.,  the
cosmological constant problem and the coincidence problem.  Thus,
some scalar field models, such as quintessence~\cite{Wetterich1988}
and phantom~\cite{Caldwell2002}, are proposed. For single scalar
field models, it has been shown that the equation of state cannot
cross the phantom divide line ($w=-1$).  So, models with a
combination of phantom and quintessence~\cite{Feng2005,Wei2005}, and
scalar field models with scalar-dependent coupling in front of
kinetic term~\cite{Nojiri2005}  as well as fluid
models~\cite{Nojiri2005b} have also been constructed to realize the
crossing of the phantom divide line, which still seems to be allowed
by recent observations~\cite{Alam2004a, Alam2004}.

Another alternative to account for the current accelerating cosmic
expansion is to modify  Einstein's general relativity theory. One
such modification is the $f(R)$ theory~\cite{Bergmann1968} (see
\cite{Felice2010} for recent reviews), where the Ricci scalar  $R$
in the Einstein-Hilbert action is generalized to  an arbitrary
function $f$ of $R$. For this theory, it has been found that the
effective equation of state can cross the phantom divide  line from
phantom phase to non-phantom one~\cite{Ali2010, Bamba2010c}.

Recently,  a new modified gravity which can also explain the
accelerating cosmic expansion~\cite{Bengochea2009}, named $f(T)$
theory, has spurred an increasing deal of interest.  The $f(T)$
theory is obtained by extending the action of teleparallel
gravity~\cite{Einstein1930} in analogy to the $f(R)$ theory, where
$T$ is the torsion scalar. An important advantage of the $f(T)$
theory is that its field equations are second order as opposed to
the fourth order equations of $f(R)$ gravity. More recently,
Linder~\cite{Linder2010} proposed some concrete $f(T)$ models (see
also Ref.~\cite{Yang2010}). We placed observational constraints on
the parameters of some of these models~\cite{Linder2010, Wu2010a,
Bengochea2010}, in particular, and analyzed the dynamical properties
of the $f(T)$ theory~\cite{ Wu2010b}, in general. A reconstruction
of the $f(T)$ theory from the background expansion history and  the
$f(T)$ theory driven by scalar fields were studied
in~\cite{Myrzakulov2010}, and the cosmological perturbations and
growth factor of matter perturbations in the $f(T)$ theory  were
investigated in Refs.~\cite{Dent2010, Zheng2010}. In addition, the
issue of local Lorentz invariance  was examined in
Refs.~\cite{Li2010, Sotiriou2010}.
 It should be noted, however, that the analysis
performed in Refs.~\cite{Wu2010a,Yang2010,Geng,Bamba2010}  indicate
that models proposed so far in the $f(T)$ theory~\cite{Linder2010,
Yang2010} behave quintessence-like or phantom-like, and thus cannot
realize the crossing of the phantom divide line for the effective
equation state, although the observational data~\cite{Alam2004a,
Alam2004} seems to indicate this crossing is still a possibility not
ruled out. So, in this paper, we propose two new $f(T)$ models which
can realize the crossing of $-1$ line, and then discuss the
constraints on model parameters from recent observations. A
remarkable feature of our models is that they realize the crossing
of  the phantom divide line from a non-phantom phase to a phantom
phase in contrast to the viable $f(R)$ models where the phantom
divide line is crossed the other way around~\cite{Bamba2010c}. It is
interesting to note that a crossing of the phantom divide from the
non-phantom phase to the phantom one is consistent with the recent
cosmological observational data~\cite{Alam2004a}. Finally, let us
note that, recently, a new model with the crossing of phantom divide
line is also proposed in \cite{Bamba2010b}.

\section{The $f(T)$ theory}
The $f(T)$ theory is obtained  by extending the action  of
teleparallel gravity to $T+f(T)$. The teleparallel theory of gravity
is built on teleparallel geometry, which uses the Weitzenb\"{o}ck
connection rather than the Levi-Civita connection. So, the spacetime
has only torsion and  is thus curvature-free.

Assuming that the universe is described by  a flat homogeneous and
isotropic Friedmann-Robertson-Walker  metric
\begin{eqnarray}
g_{\mu\nu}=\mbox{diag}(1,-a^{2}(t), -a^{2}(t), -a^{2}(t))
\;,\end{eqnarray} where $a$ is the scale factor, it has been found
in Refs.~\cite{Ferraro2007, Ferraro2008} that the torsion scalar in
the teleparallel gravity can be expressed as
\begin{eqnarray}
T=-6 H^2\;,
\end{eqnarray}
with $H=\dot{a}a^{-1}$ being the Hubble parameter.  In addition, the
modified Friedman equations have the following form
\begin{eqnarray}\label{MF}
H^2=\frac{8\pi G}{3}\rho-\frac{f}{6}+\frac{T}{3}f_{T}\;,
\end{eqnarray}
\begin{eqnarray} \label{MF2}
(H^2)'=\frac{16\pi GP-T+f-2Tf_{T}}{-4Tf_{T T}-2-2f_{T}}\;,
\end{eqnarray}
where a prime denotes a derivative with respect to $\ln a$, the
subscript $T$ represents a  derivative with respect to $T$, $\rho$
is the energy density  and $P$ is the pressure. Here we  assume that
there are  both matter and radiation components in the Universe,
thus
\begin{eqnarray}
\rho=\rho_m+\rho_r, \quad P=\frac{1}{3}\rho_r\;.
\end{eqnarray}

If we rewrite the modified Friedmann equation (Eq.~(\ref{MF})) in
the standard form as that  in general relativity, we can define an
effective dark energy, whose energy density can be expressed  as,
\begin{eqnarray}
\rho_{eff}=\frac{1}{16\pi G}(-f + 2Tf_{T})\;.
\end{eqnarray}
Here $\frac{2Tf_{T}}{f}>1$ is required in order to have a positive
value for $\rho_{eff}$. This usually gives a constraint on
physically meaningful models. Using energy conservation equation,
$\dot{\rho}_{eff}+3H(1+w_{eff}){\rho}_{eff}=0$, one can yield the
effective equation of state $w_{eff}$
\begin{eqnarray}
w_{eff}=-\frac{f/T-f_{T}+2Tf_{TT}+\frac{1}{3} \frac{8\pi G \rho_r}{ 3 H^2}(f_{T}+2T f_{T T})}{(1+f_{T}+2Tf_{TT})(f/T-2f_{T})}\;.
\end{eqnarray}
The same  expression can also be  obtained using  Eq.~(\ref{MF2}) to
define an effective pressure $p_{eff}$ and then deriving $w_{eff}$.

\section{two new $f(T)$ models}
In this section, we propose two new $f(T)$ models, labeled as Model
A and Model B, which can realize the crossing of the phantom divide
line for the effective equation of state.

  $\bullet$ Model A\begin{eqnarray}\label{M1}
         f(T)=\alpha (-T)^n \tanh\frac{T_0}{T}\;,
         \end{eqnarray}
where $\alpha$ and $n$ are two model parameters.  The requirement of
$\frac{2Tf_{T}}{f}>1$, which ensures $\rho_{eff}>0$, gives rise to
$2 n - 4x\cdot csch[2x ]>1$ with $x\equiv \frac{ T0}{T}$. Since
$4x\cdot csch[2x]\leq 2 $, $n$ must be greater than $\frac{3}{2}$.
Substituting  Eq.~(\ref{M1}) into the modified Friedmann equation,
we have
\begin{eqnarray}
\alpha=-\frac{1-\Omega_{m0}-\Omega_{r0}}{(6H_0^2)^{n-1}[2 \textrm{sech}(1)^2+(1-2n)\tanh(1)]}\;.
\end{eqnarray}
Here $\Omega_{m0}$ and $\Omega_{r0}$ are the present dimensionless density parameters of matter and radiation, respectively.

\begin{figure}[htbp]
 \centering
\includegraphics[width=6cm]{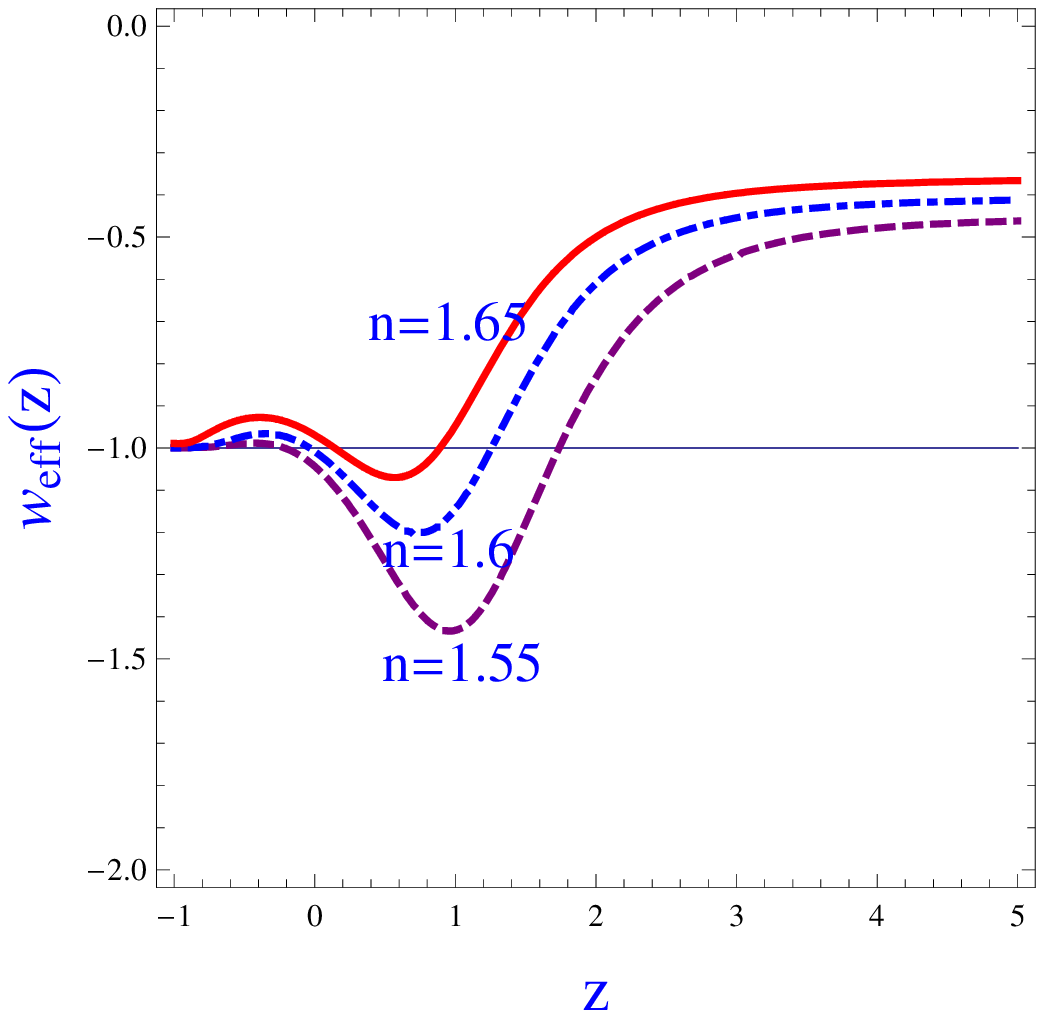}\quad \includegraphics[width=6cm]{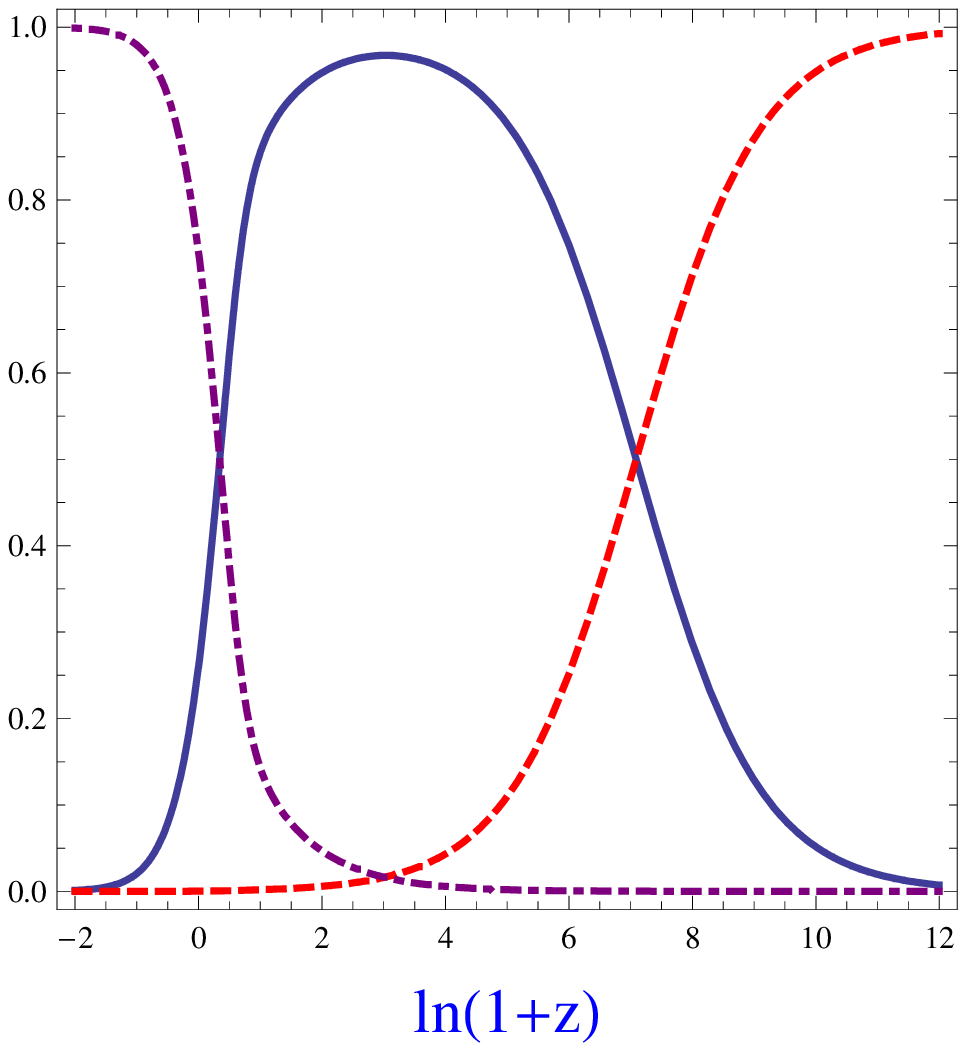}
 \caption{\label{Fig1}
 The evolutionary curves of the effective equation of state  with
 different values of $n$ and $\Omega_{m0}=0.26$ (left panel),
 and the cosmic evolution with $n=1.65$, $\Omega_{m0}=0.26$ and $\Omega_{r0}=0.26/1200$ (right panel) for Model A. In the right panel, the dot-dashed, dashed, and solid lines
 represent the evolutionary curves of the dimensionless density parameters for the effective dark energy, radiation and matter, respectively. }
 \end{figure}

\begin{figure}[htbp]
 \centering
\includegraphics[width=8cm]{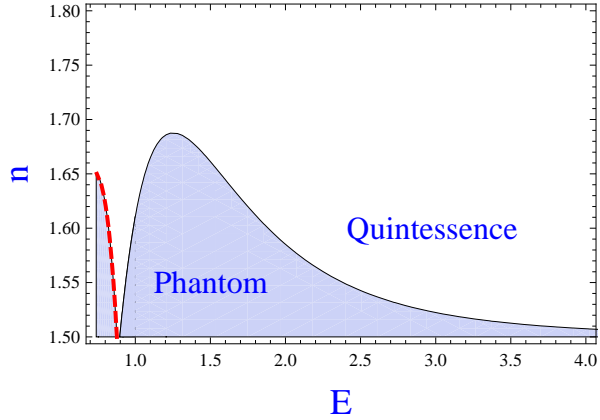}
 \caption{\label{Fig1a}
 The regions of phantom and quintessence in $n-E$ plane with $\Omega_{m0}=0.26$, where $E=H/H_0$. The red dashed line is the minimum value
 to which the universe can reach.}
 \end{figure}

In Fig.~(\ref{Fig1}),  we show the evolutionary curves of the
effective equation state with different values of $n$ (right panel)
and the cosmic evolution with $n=1.65$ (left panel). From the right
panel, one can see that the effective equation of state firstly
crosses the phantom divide line from $>-1$ (non-phantom phase) to
$<-1$ (phantom phase), and then evolves to $>-1$. So, it crosses the
$-1$ line twice. In order to illustrate why this phenomenon occurs,
we plot a figure (Fig.\ref{Fig1a}) to show the regions $w_{eff}<-1$
in $n-E$ plan with $\Omega_{m0}=0.26$, where $E=H/H_0$. From this
figure, one can see that $n$ must be smaller than a critical value,
i.e. $n<1.686$ when $\Omega_{m0}=0.26$, to render $w_{eff}$ cross
$-1$, and, once $w_{eff}$  cross the $-1$ line, it must cross it
twice. This makes the $f(T)$ models distinct from  the viable $f(R)$
models where only a crossing from phantom phase to non-phantom one
is allowed~\cite{Bamba2010c}. Finally, $w_{eff}$ approaches to $-1$,
which means that the final state of our Universe is an exponential
expansion phase. This result is consistent with what obtained in
Ref.~\cite{Wu2010b} where it has been found through the dynamical
analysis  that the Universe in the $f(T)$ theory finally enters a de
Sitter expansion phase. Furthermore, the right panel reveals that
the Universe, in this model, has a long enough period of radiation
domination to give the correct primordial nucleosynthesis and
radiation-matter equality, and    a matter dominated phase. In other
words, the usual early universe behavior can be successfully
obtained  to agree with the primordial nucleosynthesis and the
cosmic microwave background constraints.

\begin{figure}[htbp]
 \centering
\includegraphics[width=5cm]{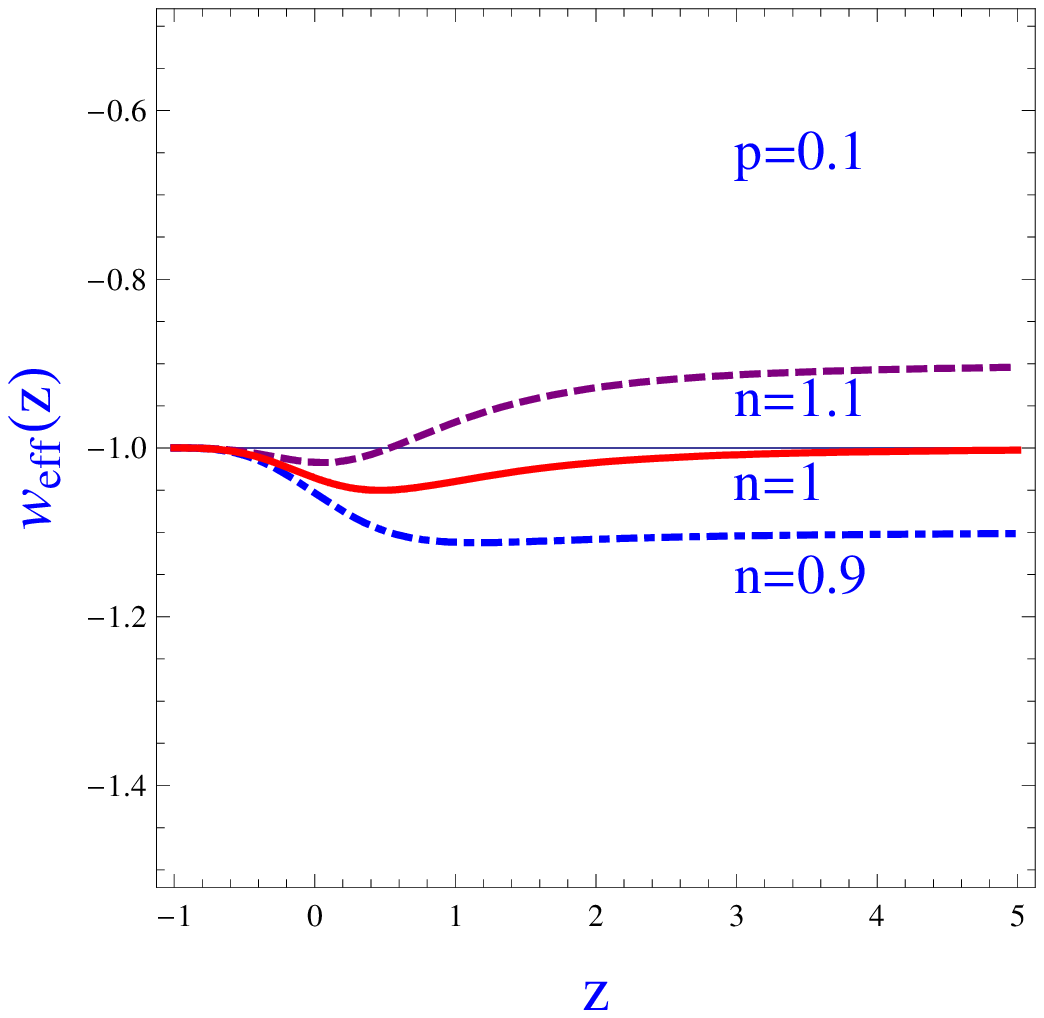}\quad \includegraphics[width=5cm]{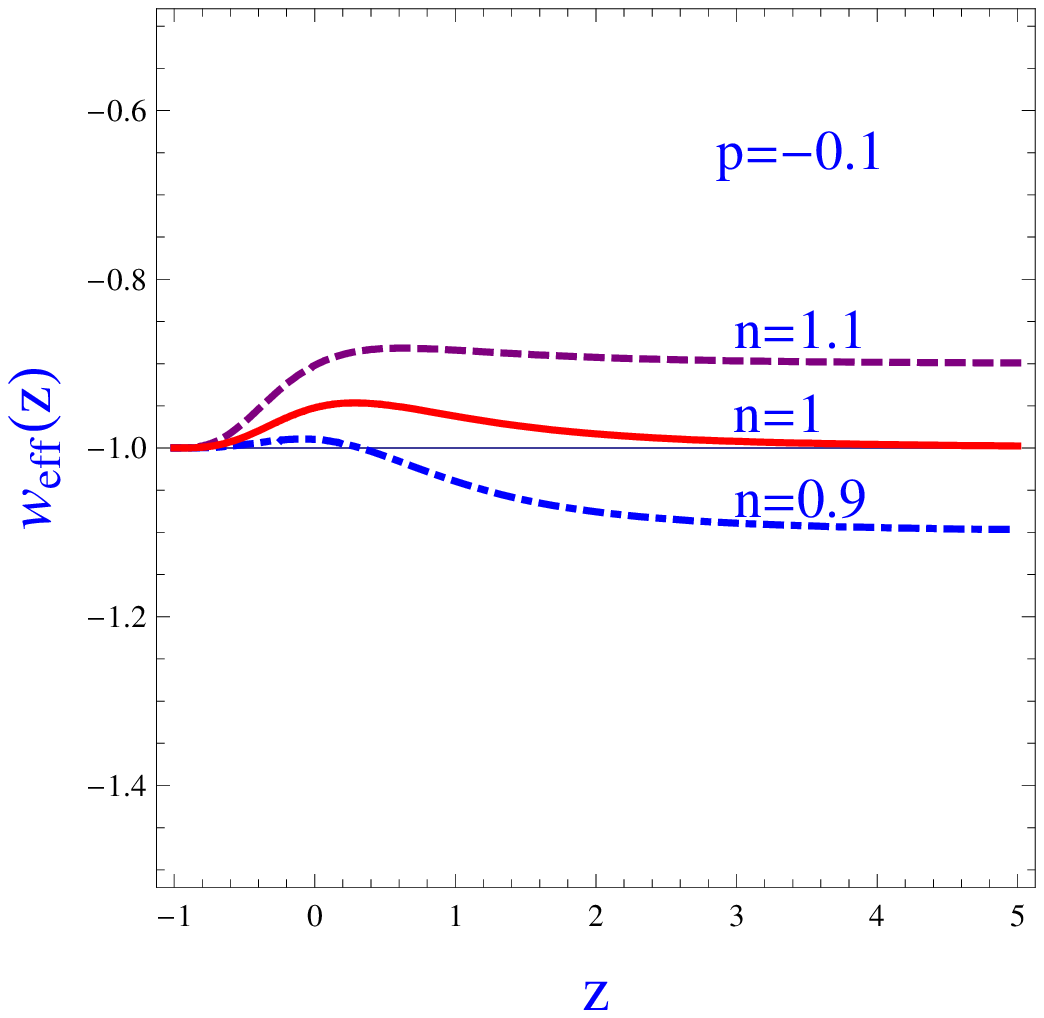}\quad \includegraphics[width=5cm]{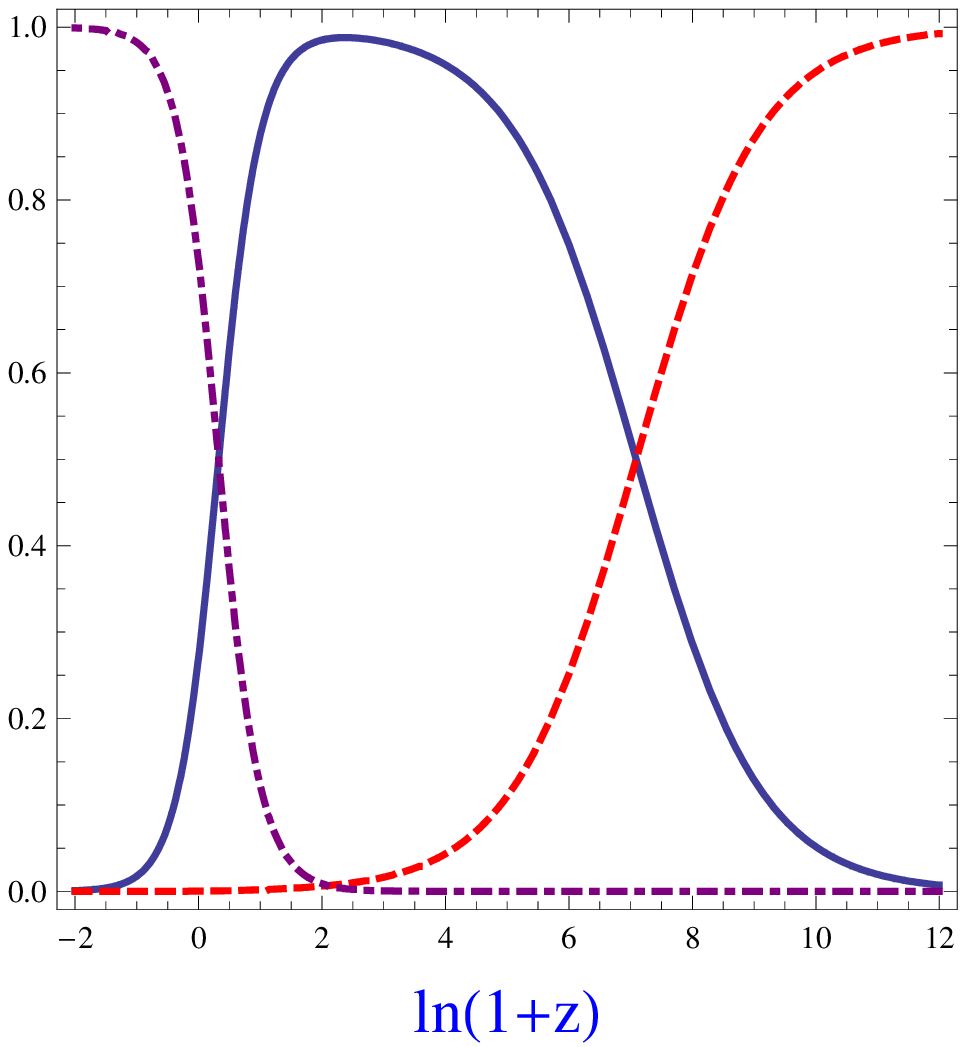}
 \caption{\label{Fig2}
 The evolutionary curves of the effective equation of state with
 different values of $p$ and $n$ and $\Omega_{m0}=0.26$ (left and middle panels),  and the cosmic evolution with $n=1.1$, $p=0.1$
 $\Omega_{m0}=0.26$ and $\Omega_{r0}=0.26/1200$(right panel) for Model B.
 In the right panel, the dot-dashed, dashed, and solid lines
 represent the evolutionary curves of the dimensionless density parameters for the effective dark energy, radiation and matter, respectively. }
 \end{figure}

$\bullet$ Model B \begin{eqnarray}
           f(T)=\alpha (-T)^n (1-e^{pT_0/T})
           \end{eqnarray}
with three model parameters $\alpha$, $n$ and $p$. From
$\frac{2Tf_{T}}{f}>1$ given by the requirement of $\rho_{eff}>0$, we
obtain that $2n-\frac{2 xe^x}{-1+e^x}>1$ with $x\equiv pT_0/T$. This
leads to $n>0.5$ since $-\frac{2 xe^x}{-1+e^x}\leq 0$. So, we now
restrict our discussion to the case of $n>\frac{1}{2}$ for Model B.
Using the modified Friedmann equation, we have
\begin{eqnarray}
\alpha=\frac{(6H_0^2)^{1-n}(1-\Omega_{m0}-\Omega_{r0})}{-1+2n+e^p(1-2n+2p)}\;.
\end{eqnarray}
 The exponential
model given by Linder~\cite{Linder2010} is a special case, the $
n=1$ case to be exact, of the present model. When $p=0$, our model
reduces to the power low model $f(T)\sim (-T)^n$, which has been
studied in detail in Refs.~\cite{Wu2010a,
Wu2010b,Linder2010,Bengochea2009,Bengochea2010}. Let us note that
when $n=1$ or $p=0$, the crossing of phantom divide line is
impossible as we have already pointed out~\cite{Wu2010b}.  This is
also confirmed  by  the $n=1$  case in Fig.~(\ref{Fig2})  in the
present paper. Fig.~(\ref{Fig2}) shows the evolutionary curves of
$w_{eff}$ (left and middle panels) and the comic evolution (right
panel) for model B. We find that, for the crossing of the $-1$ line
to occur, it is required that $p$ and $n-1$ should have the same
sign. When $p>0$ and $n>1$, $w_{eff}$ evolves from $>-1$ to $<-1$,
while, when $p<0$ and $\frac{1}{2}<n<1$, the crossing direction is
just the opposite. In addition, we also find that,  the model
behaves like quintessence when $p>0$ and $\frac{1}{2}<n<1$, and like
phantom when $p<0$ and $n>1$.  The right panel in Fig.~(\ref{Fig2})
gives the comic evolution with $n=1.1$ and $p=0.1$, from which one
can see that the usual early universe behavior can also be obtained
just as Model A.

\section{observational constraints}
Now, we discuss  the constraints on model parameters of Model A and
Model B from recent observational data, including  the Type Ia
supernovae (Sne Ia), the baryonic acoustic oscillation (BAO)
distance ratio  and the cosmic microwave background (CMB) radiation.
The Sne Ia data used in our analysis is the Union2 compilation
released by the Supernova Cosmology Project collaboration
recently~\cite{Amanullah2010}, which  consists of 557 data points
and is the largest published sample today. Using the usual method,
we constrain the theoretical model from the Sne Ia by minimizing the
$\hat{\chi}^2$ value
\begin{eqnarray}
\hat{\chi}^2_{Sne}=\sum_{i=1}^{557}\frac{[\mu_{obs}(z_i)-\mu_{th}(z_i)]^2}{\sigma_{u,i}^2}\;,
\end{eqnarray}
where  $\sigma_{\mu,i}^2$ are the errors due to the flux
uncertainties, intrinsic dispersion of Sne Ia absolute magnitude and
peculiar velocity dispersion. $\mu_{obs}$ is the observed distance
moduli and $\mu_{th}$ is the  corresponding theoretical one, which
is defined as
\begin{eqnarray}
\mu_{th}=5 \log_{10}D_L-\mu_0\;.
\end{eqnarray}
Here $\mu_0=5\log_{10}h+42.38$ with $h=H_0/100 km/s/Mpc$, and $D_L$
is the luminosity distance,
\begin{eqnarray}
D_L\equiv(1+z)\int_0^z\frac{dz'}{E(z')}\;,
\end{eqnarray}
with $E(z)\equiv H(z)/H_0$. In order to marginalize the  nuisance
parameter $\mu_0$ (or $h$), following the approach given in
Ref.~\cite{Nesseris2005}, we expand $\hat{\chi}^2_{Sne}$ to
$\hat{\chi}^2_{Sne}(\mu_0)=A \mu_0^2-2B\mu_0+C$ with
$A=\sum1/\sigma_{u,i}^2$, $B=\sum[\mu_{obs}(z_i)-5
\log_{10}D_L]/\sigma_{u,i}^2$ and $C=\sum[\mu_{obs}(z_i)-5
\log_{10}D_L]^2/\sigma_{u,i}^2$, and find that $\hat{\chi}^2_{Sne}$
has a minimum value at $\mu_0=B/A$, which is given by
\begin{eqnarray} \chi^2_{Sne}=C-\frac{B^2}{A}\;.\end{eqnarray}
Thus, we can minimize ${\chi}^2_{Sne}$ instead of
$\hat{\chi}^2_{Sne}$ to obtain  constraints from Sne Ia.

For the BAO data,   the BAO distance ratio  at $z=0.20$ and $z=0.35$
from the joint analysis of the 2dF Galaxy Redsihft Survey and SDSS
data~\cite{bao2} is used. This distance ratio
\begin{equation}
\frac{D_V(z=0.35)}{D_V(z=0.20)}=1.736\pm 0.065
\end{equation}
is a relatively model independent quantity with $D_V(z)$ defined as
\begin{equation}
D_V(z_{BAO})=\bigg[\frac{z_{BAO}}{H(z_{BAO})}\bigg(\int_0^{z_{BAO}}\frac{dz}{H(z)}\bigg)^2\bigg]^{1/3}.
\end{equation}
So, the constraint from BAO can be obtained  by performing the
following  $\chi^2$ statistics
\begin{equation}
\chi_{BAO}^2=\frac{[D_V(z=0.35)/D_V(z=0.20)-1.736]^2}{0.065^2}.
\end{equation}

Finally, we add  the CMB data in our analysis. Since  the CMB shift
parameter $R$~\cite{Wang2006, Bond1997} contains the main
information of the observations from the CMB, it is used to
constrain the theoretical models by minimizing
\begin{eqnarray}
\chi^2_{CMB}=\frac{[R-R_{obs}]^2}{\sigma_R^2}\;,
\end{eqnarray}
where $R_{obs}=1.725\pm 0.018$~\cite{Komatsu2010}, which is given by
the WMAP7 data, and  its corresponding theoretical value is defined
as
\begin{eqnarray}
R\equiv \Omega_{m0}^{1/2}\int_0^{z_{CMB}}\frac{dz'}{E(z')}\;,
\end{eqnarray}
with $z_{CMB}=1091.3$.

\begin{figure}[htbp]
 \centering
\includegraphics[width=6cm]{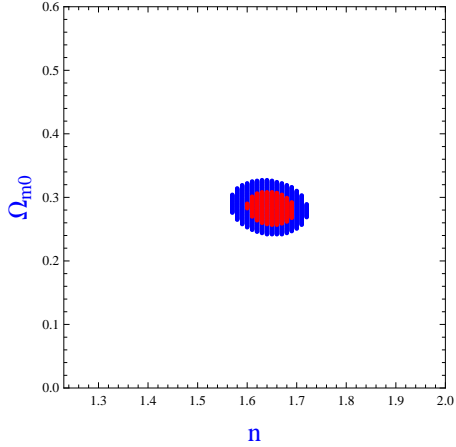}
 \caption{\label{Fig3}
 The constraint on $\Omega_{m0}$ and $n$ at the $68.3\%$ and $95.4\%$ confidence levels for Model A from Sne Ia+BAO+CMB.}
 \end{figure}

The constraints from a combination of Sne Ia, BAO and CMB can be
obtained by calculating $\chi^2_{Sne}+\chi^2_{BAO}+\chi^2_{CMB}$. We
find that, for Model A, the best fit values occur at
$\Omega_{m0}=0.282$ and $n=1.65$ with $\chi^2_{Min}=543.948$. The
contour diagrams at the $68.3\%$ and $95.4\%$ confidence levels are
given in Fig.~(\ref{Fig3}). From this figure and Fig.~(\ref{Fig1}),
we conclude that the observation favors a crossing of phantom divide
line.

\begin{figure}[htbp]
 \centering
\includegraphics[width=6cm]{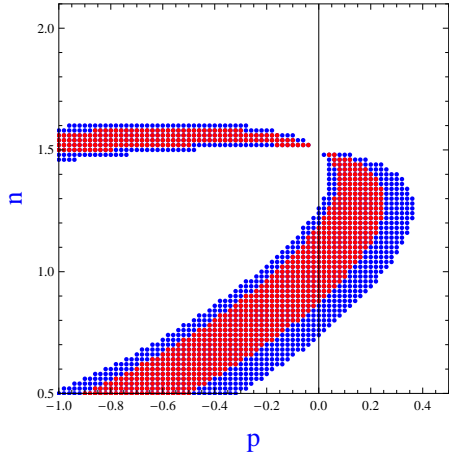}
 \caption{\label{Fig4}
 The constraint on $p$ and $n$ for Model B  with $\Omega_{m0}=0.267$ at the $68.3\%$ and $95.4\%$ confidence levels from Sne Ia+BAO+CMB.
$n>\frac{1}{2}$ given by the requirement of $\rho_{eff}>0$ is
considered. }
 \end{figure}

For Model B, the best fit values of model parameters are
$\Omega_{m0}=0.267$, $p=0.02$ and $n=1.08$ with
$\chi^2_{Min}=544.213$. It is easy to see that the best fit value
favors a crossing of the phantom divide line from $>-1$ (non-phantom
phase) to $<-1$ (phantom phase). This is consistent with the recent
observational data~\cite{Alam2004a} but is opposite to what was
found in viable $f(R)$ models~\cite{Bamba2010c}. Fig.~(\ref{Fig4})
gives the constraints in the $n-p$ plane with $\Omega_{m0}=0.267$ at
the $68.3\%$ and $95.4\%$ confidence levels, and in this figure
$n>\frac{1}{2}$ given by the requirement of $\rho_{eff}>0$ has been
taken into consideration. From Figs.~(\ref{Fig2}, \ref{Fig4}), one
can see that all possible behaviors of $w_{eff}$ shown in the left
and middle panels of Fig.~(\ref{Fig2}) are allowed by observations.

Now we consider the constraints on the $\Lambda$CDM model. The best
fit result is $\Omega_{m0}=0.270$ with $\chi^2_{Min}=544.403$. This
$\chi^2_{Min}$ is slightly larger than that obtained in the above
two $f(T)$ models. With the $\chi^2_{Min}/dof$ (dof: degree of
freedom) criterion, the $\Lambda$CDM is slightly favored  by
observations.

\section{Conclusion}
The $f(T)$ theory is a new modified gravity, obtained by extending
the teleparallel gravity, to account for the current accelerating
cosmic expansion without the need of dark energy.  In this paper, we
have proposed two new $f(T)$ models in which the crossing of the
phantom divide line is possible. A remarkable feature of the our
models is that they realize the
 crossing of  the phantom divide line from a non-phantom phase to a phantom
 phase in contrast to the viable $f(R)$ models where the phantom divide
 line is crossed the other way around~\cite{Bamba2010c}. It is interesting to note that a crossing of the phantom divide from
 the non-phantom phase to the phantom one  is consistent with the recent cosmological observational
 data~\cite{Alam2004a}. By studying the evolutionary
curves of $w_{eff}$, we find that  $w_{eff}$ can cross the $-1$ line
in both models and it is crossed twice in Model A. Furthermore, we
also find that both models can produce the usual early universe
behaviors in the sense that they both allow a long enough period of
radiation domination and a matter dominated phase to agree with the
primordial nucleosynthesis and the cosmic microwave background
constraints. We have also discussed the constraints on model
parameters from recent observations including Sne Ia, BAO and CMB.
Our results show that observations favor a crossing of the $-1$ line
for Model A, whereas, for Model B, all possible evolutions for
$w_{eff}$ given in Fig.~(\ref{Fig2}) are allowed, although the best
fit result favors the crossing. With the $\chi^2_{Min}/dof$ (dof:
degree of freedom) criterion, we find that the $\Lambda$CDM is still
favored slightly by observations.

\begin{acknowledgments}

We thank the anonymous referee for helpful comments and suggestions.
This work was supported in part by the National Natural Science
Foundation of China under Grants Nos. 10935013 and 11075083,
Zhejiang Provincial Natural Science Foundation of China under Grant
No. Z6100077, the FANEDD under Grant No. 200922, the National Basic
Research Program of China
under Grant No. 2010CB832803, the NCET under Grant No. 09-0144, 
and K.C. Wong Magna Fund in Ningbo
University.

\end{acknowledgments}


\begin{thebibliography}{99}
\bibitem{Riess1998}  A. G.  Riess, et al.,  Astron. J. {\bf 116}, 1009 (1998);  S. Perlmutter, et al., Astrophys. J. {\bf 517}, 565 (1999).
\bibitem{Spergel2003} D. N. Spergel, et al., ApJS, {\bf  148}, 175 (2003);
                      D. N. Spergel, et al., ApJS, {\bf 170}, 377S (2007).
\bibitem{Tegmark2004} M. Tegmark, et al., Phys. Rev. D {\bf 69},  103501 (2004).
\bibitem{Eisenstein2005} D. J. Eisenstein, et al., Astrophys. J. {\bf 633}, 560 (2005).

\bibitem{Weinberg1989} S. Weinberg, Rev. Mod. Phys. {\bf 61}, 1  (1989);
                      V. Sahni  and A. Starobinsky,  Int. J. Mod. Phys. D {\bf 9},  373 (2000);
                      P. J. E. Peebles and  B. Ratra,  Rev. Mod. Phys. {\bf 75}, 559 (2003);
                      T.  Padmanabhan,  Phys. Rep. {\bf 380}, 235 (2003).
\bibitem{Wetterich1988}C. Wetterich, Nucl. Phys. B {\bf 302}, 668  (1988);
                       B. Ratra  and P.  J. E. Peebles,  Phys. Rev. D {\bf 37}, 3406 (1988);
                       R. R. Caldwell, R. Dave and P. J. Steinhardt,  Phys. Rev. Lett. {\bf 80},  1582 (1998).

\bibitem{Caldwell2002} R. R. Caldwell,  Phys. Lett. B {\bf 545},  23 (2002);
                       R. R. Caldwell, M. Kamionkowski and N. N. Weinberg,  Phys. Rev. Lett. {\bf 91}, 071301 (2003);
                       S. Nesseris  and L. Perivolaropoulos,  Phys. Rev. D {\bf 70},  123529 (2004);
                       S. Nojiri and S. D. Odintsov, Phys. Lett. B {\bf 562}, 147  (2003);
                       S. Nojiri, S. D. Odintsov and S. Tsujikawa, Phys. Rev. D {\bf 71}, 063004 (2005);
                       P. Wu and H. Yu,  J. Cosmol. Astropart. Phys. {\bf 05}, 008 (2006);
                       P. Wu and H. Yu,  Nucl. Phys. B {\bf 727},  355 (2005);
                       R. Gannouji, D. Polarski, A. Ranquet, A. A. Starobinsky, JCAP {\bf 0609}, 016 (2006);
                       C. J. Feng, X. Li, E. N. Saridakis, Phys. Rev. D {\bf 82}, 023526 (2010).



\bibitem{Feng2005}  B. Feng, X. Wang and X. Zhang,  Phys. Lett. B {\bf 607},  35 (2005);
                    E. Elizalde, S. Nojiri, S. D. Odintsov, Phys. Rev. D {\bf 70}, 043539 (2004);
                    Z.  Guo, Y. Piao, X. Zhang and Y. Zhang, Phys. Lett. B {\bf 608},  177 (2005);
                    P.  Wu  and H. Yu, Int. J. Mod. Phys. D {\bf 14},  1873 (2005);
                    Y. Cai, H. Li, Y. Piao and X. Zhang, Phys. Lett. B {\bf 646},  141  (2007);
                     L. P. Chimento, M. Forte, R. Lazkoz, M. G. Richarte, Phys. Rev. D {\bf 79}, 043502 (2009);
                    Y. Cai, E. N. Saridakis, M.R. Setare, J. Xia, Phys. Rep. {\bf 493},  1-60 (2010).



\bibitem{Wei2005} H. Wei, R. G. Cai  and D. Zeng,  Class. Quantum Grav. {\bf 22}, 3189 (2005);
           H. Wei, R. G. Cai,  Phys. Rev. D {\bf 72}, 123507 (2005);
           M. Alimohammadi and H. Mohseni Sadjadi,  Phys. Rev. D {\bf 73},  083527
           (2006);
          W. Zhao  and Y. Zhang, Phys. Rev. D {\bf 73},  123509 (2006);
          H. Wei, N. N. Tang and S. N. Zhang,  Phys. Rev. D {\bf 75}, 043009
          (2007).
\bibitem{Nojiri2005}S. Nojiri and S. D. Odintsov, Gen. Rel. Grav. {\bf 38}, 1285 (2006);
                    S. Capozziello, S. Nojiri, S. D. Odintsov,  Phys. Lett. B {\bf 632}, 597 (2006).

\bibitem{Nojiri2005b}S. Nojiri, S. D. Odintsov, Phys. Rev. D {\bf 72}, 023003 (2005).


\bibitem{Alam2004a}
     U. Alam, V. Sahni, A. A. Starobinsky, J. Cosmol. Astropart. P. {\bf 0406}, 008 (2004);
     S. Nesseris and L. Perivolaropoulos, J. Cosmol. Astropart. P. {\bf 0701}, 018 (2007);
     P. Wu and H. Yu, Phys. Lett. B {\bf 643}, 315 (2006);
     U. Alam, V. Sahni and A. A. Starobinsky, J. Cosmol. Astropart. P. {\bf 0702}, 011 (2007);
     H. K. Jassal, J. S. Bagla and T. Padmanabhan, Mon. Not. Roy.  Astron.  Soc.  {\bf 405}, 2639 (2010).


\bibitem{Alam2004}   U. Alam, V. Sahni, T. D. Saini, A. A. Starobinsky, Mon. Not. Roy. Astron. Soc. {\bf 354}, 275 (2004);
     Y. Wang and P. Mukherjee, Astrophys. J. {\bf 606}, 654 (2004);
     R. Lazkoz, S. Nesseris and L. Perivolaropoulos, J. Cosmol. Astropart. P. {\bf 0511}, 010 (2005);
     Y. G. Gong and A. Wang, Phys. Rev. D {\bf 75}, 043520 (2007);
     Y. G. Gong, R. G. Cai, Y. Chen and Z. H. Zhu, J. Cosmol. Astropart. Phys. {\bf 01},   019 (2010);
     H. Zhang,  arXiv:0909.3013.

\bibitem{Bergmann1968}P. G. Bergmann, Int. J. Theor. Phys. {\bf 1}, 25 (1968);
                      T. V. Ruzmaikina and A. A. Ruzmaikin, Zh. Eksp. Teor. Fiz. {\bf 57}, 680 (1969) [Sov. Phys. - JETP 30, 372 (1970)];
       H. A. Buchdahl, Mon. Not. Roy. Astron. Soc. {\bf 150}, 1 (1970);
       A. A. Starobinsky, Phys. Lett. B {\bf 91}, 99 (1980);
       S. Capozziello, Int. J. Mod. Phys. D {\bf 11}, 483, (2002);
       S. Capozziello, S. Carloni and A. Troisi, Recent Res. Dev. Astron. Astrophys. {\bf 1}, 625 (2003);
       S. Capozziello, V. F. Cardone, S. Carloni and A. Troisi, Int. J. Mod. Phys. D {\bf  12}, 1969 (2003);
       S. M. Carroll, V. Duvvuri, M. Trodden and M. S. Turner, Phys. Rev. D {\bf 70}, 043528 (2004);
       S. Nojiri and S. D. Odintsov,  Phys. Rev. D {\bf 68}, 123512 (2003).




\bibitem{Felice2010}S. Nojiri and S. D. Odintsov, Int. J. Geom. Meth. Mod. Phys. {\bf 4}, 115 (2007);
                    T. P. Sotiriou and V. Faraoni, Rev. Mod. Phys. {\bf 82}, 451 (2010);
                    A. De Felice and S. Tsujikawa, Living Rev. Rel. {\bf 13},  3  (2010);
                    S. Nojiri and S. D. Odintsov,  arXiv:1011.0544.

\bibitem{Ali2010} A. Ali, R. Gannouji, M. Sami, A. A. Sen, Phys. Rev. D {\bf 81}, 104029 (2010).
\bibitem{Bamba2010c}K. Bamba, C. Q. Geng and C. C. Lee, JCAP 1008, 021 (2010);
                    K. Bamba, C. Q. Geng and C. C. Lee, arXiv:1007.0482.

\bibitem{Bengochea2009} G. R. Bengochea and R. Ferraro, Phys. Rev. D {\bf 79}, 124019 (2009).

\bibitem{Einstein1930}A. Einstein, Sitzungsber. Preuss. Akad. Wiss. Phys. Math. Kl., 217 (1928); 401 (1930);
                      A. Einstein, Math. Ann. {\bf 102}, 685 (1930);
                      K. Hayashi and T. Shirafuji, Phys. Rev. D {\bf 19}, 3524 (1979); {\bf 24}, 3312 (1981).

\bibitem{Linder2010}    E. V. Linder, Phys. Rev. D {\bf 81}, 127301 (2010).
\bibitem{Yang2010} R. Yang,  arXiv:1007.3571.

\bibitem{Wu2010a} P. Wu and H. Yu,   Phys. Lett. B {\bf 693}, 415 (2010). 
\bibitem{Bengochea2010} G. R. Bengochea, arXiv:1008.3188.
\bibitem{Wu2010b} P. Wu and H. Yu, Phys. Lett. B {\bf 692}, 176 (2010). 



\bibitem{Myrzakulov2010} R. Myrzakulov, arXiv:1006.1120;
                 K. K. Yerzhanov, Sh. R. Myrzakul, I. I. Kulnazarov, R. Myrzakulov, arXiv:1006.3879;
                 R. Myrzakulov,  arXiv:1008.4486;
                 K. Karami, A. Abdolmaleki,  arXiv:1009.2459;
                 K. Karami, A. Abdolmaleki,  arXiv:1009.3587;
                 P. Yu. Tsyba, I. I. Kulnazarov, K. K. Yerzhanov, R. Myrzakulov, arXiv:1008.0779.
\bibitem{Dent2010} S. Chen, J. B. Dent, S. Dutta, E. N. Saridakis, arXiv:1008.1250;
                   J. B. Dent, S. Dutta, E. N. Saridakis, arXiv:1010.2215;
\bibitem{Zheng2010} R. Zheng and Q. Huang, arXiv:1010.3512.

\bibitem{Bamba2010}K. Bamba, C. Q. Geng, C. C. Lee, L. W. Luo, arXiv:1011.0508.


\bibitem{Sotiriou2010} T. P. Sotiriou, B. Li, J. D. Barrow, arXiv:1012.4039

\bibitem{Li2010} B. Li, T. P. Sotiriou, J. D. Barrow,  arXiv:1010.1041



\bibitem{Geng}    C. Q. Geng, Talk given at the ITPC Workshop on Dark Energy and  Dark Matter at Weihai, China, August, 2010.
\bibitem{Bamba2010b}K. Bamba, C. Q. Geng, C. C. Lee, arXiv:1008.4036.

\bibitem{Ferraro2007}   R. Ferraro and F. Fiorini, Phys. Rev. D {\bf 75}, 084031 (2007).
\bibitem{Ferraro2008}   R. Ferraro and F. Fiorini, Phys. Rev. D {\bf 78}, 124019 (2008).


\bibitem{Amanullah2010}  R. Amanullah, et al.,  arXiv:1004.1711.
\bibitem{Nesseris2005}S. Nesseris and L. Perivolaropoulos, Phys. Rev. D {\bf 72}, 123519 (2005).

\bibitem{bao2}
       B. A. Reid, {\it et al.},  Mon. Not. Roy. Astron. Soc. {\bf 404}, 60 (2010);
        W. J. Percival, {\it et al.}, Mon. Not. Roy. Astron. Soc. {\bf 401}, 2148 (2010).

\bibitem{Wang2006}    Y. Wang and P. Mukherjee, Astrophys. J. {\bf 650}, 1 (2006).
\bibitem{Bond1997}    J. R. Bond, G. Efstathiou and M. Tegmark, Mon. Not. Roy. Astron. Soc. {\bf 291}, L33 (1997)
\bibitem{Komatsu2010} E. Komatsu et al., arXiv:1001.4538.


\end{thebibliography}
\end{document}